\documentclass[aps,a4paper,superscriptaddress,showpacs,preprintnumbers,amsmath,amssymb]{revtex4}

\usepackage{ulem}

\usepackage{psfrag} \usepackage{graphicx} \usepackage{dcolumn}
\usepackage{color} \usepackage{latexsym,amsfonts} \usepackage{bm}
\usepackage{amssymb}
\begin{document}

\title{Neutron Dark Matter Decays\\ and Correlation Coefficients of
  Neutron $\beta^-$--Decays}

\author{A. N. Ivanov}\email{ivanov@kph.tuwien.ac.at}
\affiliation{Atominstitut, Technische Universit\"at Wien, Stadionallee
  2, A-1020 Wien, Austria}
\author{R.~H\"ollwieser}\email{roman.hoellwieser@gmail.com}
\affiliation{Atominstitut, Technische Universit\"at Wien, Stadionallee
  2, A-1020 Wien, Austria}\affiliation{Department of Physics,
  Bergische Universit\"at Wuppertal, Gaussstr. 20, D-42119 Wuppertal,
  Germany} \author{N. I. Troitskaya}\email{natroitskaya@yandex.ru}
\affiliation{Atominstitut, Technische Universit\"at Wien, Stadionallee
  2, A-1020 Wien, Austria}
\author{M. Wellenzohn}\email{max.wellenzohn@gmail.com}
\affiliation{Atominstitut, Technische Universit\"at Wien, Stadionallee
  2, A-1020 Wien, Austria} \affiliation{FH Campus Wien, University of
  Applied Sciences, Favoritenstra\ss e 226, 1100 Wien, Austria}
\author{Ya. A. Berdnikov}\email{berdnikov@spbstu.ru}\affiliation{Peter
  the Great St. Petersburg Polytechnic University, Polytechnicheskaya
  29, 195251, Russian Federation}

\date{\today}

\begin{abstract}
As we have pointed out in (arXiv:1806.10107 [hep-ph]), the existence
of neutron dark matter decay modes $n \to \chi + anything$, where
$\chi$ is a dark matter fermion, for the solution of the neutron
lifetime problem changes priorities and demands to describe the
neutron lifetime $\tau_n = 888.0(2.0)\,{\rm s}$, measured in beam
experiments and defined by the decay modes $n \to p + anything$, in
the Standard Model (SM). The latter requires the axial coupling
constant $\lambda$ to be equal to $\lambda = - 1.2690$
(arXiv:1806.10107 [hep-ph]). Since such an axial coupling constant is
excluded by experimental data reported by the PERKEO II and UCNA
Collaborations, the neutron lifetime $\tau_n = 888.0(2.0)\,{\rm s}$
can be explained only by virtue of interactions beyond the SM, namely,
by the Fierz interference term of order $b \sim - 10^{-2}$ dependent
on scalar and tensor coupling constants. We give a complete analysis
of all correlation coefficients of the neutron $\beta^-$--decays with
polarized neutron, taking into account the contributions of scalar and
tensor interactions beyond the SM with the Fierz interference term $b
\sim - 10^{-2}$. We show that the obtained results agree well with
contemporary experimental data that does not prevent the neutron with
the rate of the decay modes $n \to p + anything$, measured in beam
experiments, to have dark matter decay modes $n \to \chi + anything$.
\end{abstract} 

\maketitle

\section{Introduction}
\label{sec:introduction}

Recently Fornal and Grinstein \cite{Fornal2018,Fornal2018a} have
proposed to explain the neutron lifetime anomaly, related to a
discrepancy between experimental values of the neutron lifetime
measured in bottle and beam experiments, through contributions of the
neutron dark matter decay modes $n \to \chi + \gamma$, $n \to \chi +
\phi$ and $n \to \chi +\gamma^* \to \chi + e^- + e^+$, where $\chi$
and $\phi$ are dark matter fermion and scalar boson, respectively,
$\gamma$ and $\gamma^*$ are real and virtual photons, and $(e^- e^+)$
is the electron--positron pair. However, according to recent
experimental data \cite{Tang2018,Sun2018}, the decay modes $n \to \chi
+ \gamma$ and $n \to \chi + e^- + e^+$ are suppressed.  In turn, the
neutron decay mode $n \to \chi + \phi$ has not been investigated
experimentally. In \cite{Ivanov2018d} the experimental data on the
decay mode $n \to \chi + e^- + e^+$ in \cite{Sun2018} have been
interpreted as follows. An unobservability of the decay mode $n \to
\chi + e^- + e^+$, which is not mediated by a virtual photon, may also
mean that the production of the electron--positron pair in such a
decay is below the reaction threshold, i.e. a mass $m_{\chi}$ of dark
matter fermions $\chi$ obeys the constraint $ m_{\chi} > m_n - 2 m_e$,
where $m_n$ and $m_e$ are masses of the neutron and electron
(positron), respectively. Then, we have proposed that the neutron
lifetime anomaly can be explained by the decay mode $n \to \chi +
\nu_e + \bar{\nu}_e$, where $(\nu_e \bar{\nu}_e)$ is a
neutrino--antineutrino pair \cite{Ivanov2018d}. Since neutrino $\nu_e$
and electron $e^-$ belong to the same doublet in the Standard
Electroweak Model (SEM) \cite{PDG2018}, neutrino--antineutrino $(\nu_e
\bar{\nu}_e)$ pairs couple to the neutron--dark matter current with
the same strength as electron--positron $(e^-e^+)$ pairs
\cite{Ivanov2018d}. For the UV completion of the effective interaction
($n \chi \ell\bar{\ell}$), where $\ell(\bar{\ell})$ is electron
(positron) or neutrino(antineutrino), we have proposed a gauge
invariant quantum field theory model with $SU_L(2)\times U_Y(1) \times
U_R'(1)\times U''_L(1)$ gauge symmetry, where $Y$ is the weak
hypercharge \cite{PDG2018}. Such a quantum field theory model contains
the sector of the SEM (or the Standard Model (SM) sector)
\cite{PDG2018} with $SU_L(2)\times U_Y(1)$ gauge symmetry and the dark
matter sector with $U'_R(1) \times U''_L(1)$ gauge symmetry. In the
physical phase the dark matter sectors with $U'_R(1)$ and $U''_L(1)$
symmetries are responsible for the UV completion of the effective
interaction ($n \chi \ell\bar{\ell}$) \cite{Ivanov2018d} and
interference of the dark matter into dynamics of neutron stars
\cite{McKeen2018,Motta2018,Baym2018,Cline2018}, respectively. The dark
matter sector with $U''_L(1)$ gauge symmetry we have constructed in
analogue with scenario proposed by Cline and Cornell
\cite{Cline2018}. This means that dark matter fermions with mass
$m_{\chi} < m_n$ couple to a very light dark matter spin--1 boson
$Z''$ providing a necessary repulsion between dark matter fermions in
order to give a possibility for neutron stars to reach masses of about
$2 M_{\odot}$ \cite{Demorest2010}, where $M_{\odot}$ is the mass of
the Sun \cite{PDG2018}. We have shown that in the physical phase the
predictions of the dark matter sector with $U'_R(1)$ gauge symmetry do
not contradict constraints on i) dark matter production in ATLAS
experiments at the LHC, ii) the cross section for the low--energy dark
matter fermion--electron scattering ($\chi + e^- \to \chi + e^-$)
\cite{Essig2017}, and iii) the branching ratio of the Higgs--boson
decay $H^0 \to Z + Z'$ \cite{Curtin2014}, where $H^0$, $Z$ and $Z'$
are the Higgs--boson with mass $M_{H^0} = 125\,{\rm GeV}$
\cite{PDG2018}, the electroweak $Z$--boson \cite{PDG2018} and the dark
matter spin--1 boson $Z'$ of the dark matter sector with $U'_R(1)$
gauge symmetry \cite{Ivanov2018d}. We have also proposed that the
reactions $n \to \chi + \nu_e + \bar{\nu}_e$, $n + n \to \chi + \chi$,
$n + n \to \chi + \chi + \nu_e + \bar{\nu}_e$ and $\chi + \chi \to n +
n$, allowed in our model, can serve as URCA processes for the neutron
star cooling \cite{Gamow1941,Friman1979,Hansel1995}. In addition we
have discussed a possible quark level formulation of our model with
$SU_L(2) \times U_Y(1) \times U'_R(1) \times U''_L(1)$ gauge symmetry.

Having assumed that the results of the experimental data
\cite{Tang2018,Sun2018} can be also interpreted as a production of
electron--positron pairs below reaction threshold of the decay mode $n
\to \chi + e^- + e^+$, we have proposed to search for traces of dark
matter fermions induced by the $n\chi\,e^- e^+$ interaction in the
low--energy electron--neutron inelastic scattering $e^- + n \to \chi +
e^-$. Such a reaction can be compared experimentally with low--energy
electron--neutron elastic scattering $e^- + n \to n + e^-$
\cite{Schwinger1948}--\cite{Yennie1957}. The differential cross
section for the reaction $e^- + n \to \chi + e^-$ possesses the
following properties: i) it is inversely proportional to a velocity of
incoming electrons, ii) it is isotropic relative to outgoing electrons
and iii) momenta of outgoing electrons are much larger than momenta of
incoming electrons. Because of these properties the differential cross
section for the reaction $e^- + n \to \chi + e^-$ can be in principle
distinguished above the background of the elastic electron--neutron
scattering $e^- + n \to n + e^-$. In order to have more processes with
particles of the SM in the initial and final states allowing to search
dark matter in terrestrial laboratories we have proposed to search
dark matter fermions by means of the electrodisintegration of the
deuteron into dark matter fermions and protons $e^- + d \to \chi + p +
e^-$ close to threshold \cite{Ivanov2018e}, induced by the
electron--neutron inelastic scattering $e^- + n \to \chi + e^-$ with
energies of incoming electrons larger than the deuteron binding
energy, which is of about $|\varepsilon_d| \sim 2\,{\rm MeV}$.  We
have calculated the triple--differential cross section for the
reaction $e^- + d \to \chi + p + e^-$ close to threshold, and proposed
to detect dark matter fermions from the electrodisintegration of the
deuteron $e^- + d \to \chi + p + e^-$ above the background $e^- + d
\to n + p + e^-$ by detecting outgoing electrons, protons and neutrons
in coincidence. A missing of neutron signals at simultaneously
detected signals of protons and outgoing electrons should testify an
observation of dark matter fermions in the final state of the
electrodisintegration of the deuteron close to threshold.

As has been pointed out in \cite{Ivanov2018d}, the acceptance of
existence of the neutron dark matter decay modes $n \to \chi + {\rm
  anything}$ is not innocent and demands to pay the following
price. Indeed, the neutron lifetime time $\tau_n = 879.6(1.1)\,{\rm
  s}$, calculated in the SM \cite{Ivanov2013} for the axial coupling
constant $\lambda = - 1.2750(9)$ \cite{Abele2008} by taking into
account the complete set of corrections of order $10^{-3}$, caused by
the weak magnetism and proton recoil, taken to next--to--leafing order
in the large nucleon mass $M$ expansion of order $O(E_e/M)$, where
$E_e$ is the electron energy, and radiative corrections of order
$O(\alpha/\pi)$, where $\alpha$ is the fine--structure constant
\cite{PDG2018}, agrees well with the world averaged lifetime of the
neutron $\tau_n = 880.1(1.0)\,{\rm s}$ \cite{PDG2018}, and the neutron
lifetime $\tau_n = 879.6(6)$, averaged over the experimental values
measured in bottle experiments \cite{Mampe1993}--\cite{Arzumanov2015}
included in the Particle Date Group (PDG) \cite{PDG2018}. It agrees
also with the value $\tau_n = 879.4(6)\,{\rm s}$ and the axial
coupling constant $\lambda = - 1.2755(11)$, obtained by Czarnecki {\it
  et al.}  \cite{Czarnecki2018} by means of a global analysis of the
experimental data on the neutron lifetime and axial coupling
constant. At first glimpse such an agreement rules out fully any dark
matter decay mode $n \to \chi + anything$ of the neutron. For a
possibility to the neutron to have any dark matter decay mode $n \to
\chi + anything$ the SM should explain the neutron lifetime $\tau_n =
888.0(2.0)\,{\rm s}$, measured in beam experiments, instead of to
explain the neutron lifetime $\tau_n = 879.6(6)\,{\rm s}$, measured in
bottle ones. As has been shown in \cite{Ivanov2018d}, using the
analytical expression for the neutron lifetime (see Eq.(41) and (42)
of Ref.\cite{Ivanov2013}) the value $\tau_n = 888.0(2.0)\,{\rm s}$ can
be fitted by the axial coupling constant equal to $\lambda = -
1.2690$. Since such a value of the axial coupling constant is ruled
out by recent experiments \cite{Abele2013}--\cite{Brown2018} and a
global analysis by Czarnecki {\it et al.} \cite{Czarnecki2018}, so the
hypothesis of the existence of the neutron dark matter decay modes
should state that the SM, including a complete set of corrections of
order $10^{-3}$ caused by the weak magnetism, proton recoil and
radiative corrections \cite{Gudkov2006} (see also \cite{Ivanov2013}),
is not able to describe correctly the rate and correlation
coefficients of the neutron decay modes $n \to p + anything$. Hence,
the theoretical description of the neutron lifetime, measured in beam
experiments, should go beyond the SM. Indeed, keeping the value of the
axial coupling constant equal to $\lambda = - 1.2750$ or so
\cite{Abele2013}--\cite{Brown2018} and having accepted the existence
of the dark matter decay modes $n \to \chi + {\rm anything}$ we have
also to accept a sufficiently large contribution of the Fierz
interference term $b$ \cite{Fierz1937}, dependent on the scalar and
tensor coupling constants of interactions beyond the SM
\cite{Fierz1937}--\cite{Ivanov2018a} (see also \cite{Gudkov2006} and
\cite{Ivanov2013}). Using the results obtained in \cite{Ivanov2013},
the neutron lifetime $\tau_n = 888.0\,{\rm s}$ can be fitted by the
axial coupling constant $\lambda = -1.2750$, the
Cabibbo--Kobayashi--Maskawa (CKM) matrix element $V_{ud} = 0.97420$
\cite{PDG2018} and the Fierz interference term $b = - 1.44 \times
10^{-2}$ calculated at the neglect of the quadratic contributions of
scalar and tensor coupling constants of interactions beyond the SM
\cite{Ivanov2018d}. Thus, in order to confirm a possibility for the
neutron to have any dark matter decay modes $n \to \chi + anything$ we
have to show that a tangible influence of the Fierz interference term
$b = - 1.44 \times 10^{-2}$ is restricted only by the rate $1/\tau_n =
1/888.0\,{\rm s^{-1}} = 1.126\times 10^{-3}\,{\rm s^{-1}}$ of the
neutron decay modes $n \to p + anything$, measured in beam
experiments, and such a term does not affect the correlation
coefficients of the electron--energy and angular distributions of the
neutron $\beta^-$--decay.

This paper is addressed to the analysis of the contributions of scalar
and tensor interactions beyond the SM to the rate of the neutron decay
modes $n \to p + anything$, measured in beam experiments, and
correlation coefficients of the neutron $\beta^-$--decay with
polarized neutron, polarized electron and unpolarized proton. We take
into account the contributions of the SM, including a complete set of
corrections of order $10^{-3}$, caused by the weak magnetism and
proton recoil, calculated to next--to--leading order in the large
nucleon mass $M$ expansion \cite{Bilenky1959,Wilkinson1982} (see also
\cite{Gudkov2006} and \cite{Ivanov2013,Ivanov2017b}), and radiative
corrections of order $O(\alpha/\pi)$, calculated to leading order in
the large nucleon mass expansion
\cite{Sirlin1967}--\cite{Czarnecki2004} (see also \cite{Gudkov2006}
and \cite{Ivanov2013,Ivanov2017b}). We search anyone solution for the
values of scalar and tensor coupling constants of interactions beyond
the SM allowing to fit the rate $1/\tau_n = 1.126\times 10^{-3}\,{\rm
  s^{-1}}$ of the neutron decay modes $n \to p + {\rm anything}$,
measured in beam experiments, and the experimental data on the
correlation coefficients of the neutron $\beta^-$--decay under
consideration. The existence of such a solution for the values of
scalar and tensor coupling constants of interactions beyond the SM
should imply an allowance for the neutron to have the dark matter
decay modes $n \to \chi + anything$.

The paper is organized as follows. In section \ref{sec:distribution}
we give the electron--energy and angular distribution of the neutron
$\beta^-$--decay with polarized neutron, polarized electron and
unpolarized proton. We write down the expressions for the correlation
coefficients including the contributions of the SM corrections of
order $10^{-3}$, caused by the weak magnetism and proton recoil to
next--to--leading order in the large nucleon mass expansion of order
$O(E_e/M)$ and radiative corrections of order $O(\alpha/\pi)$, and the
contributions of scalar and tensor interactions beyond the SM,
calculated to leading order in the large nucleon mass $M$
expansion. In section \ref{sec:lifetime} we analyse the rate $1/\tau_n
= 1.126\times 10^{-3}\,{\rm s^{-1}}$ of the neutron decay modes $n \to
p + anything$, measured in beam experiments, and fit it by the
contribution of the Fierz interference term by taking into account the
quadratic contributions of scalar and tensor interactions beyond the
SM.  In section \ref{sec:coefficient} we analyse possible solutions
for the values of scalar and tensor coupling constants. On this way
for a search of one of possible solutions we follow
\cite{Severijns2006,Ivanov2018a} and assume that the scalar and tensor
coupling constants are real, and set $\bar{C}_S = - C_S$ and $C_T = -
\bar{C}_T$. Such a solution implies also that in scalar and tensor
interactions beyond the SM the neutron and proton couple to
right--handed electron and antineutrino only. We take into account the
constraints on the scalar coupling constant $C_S$, i.e. $|C_S| =
0.0014(13)$ and $|C_S| = 0.0014(12)$ obtained by Hardy and Towner
\cite{Hardy2015} and Gonz\'alez--Alonso {\it et al.}
\cite{Severijns2018}, respectively, from the superallowed $0^+ \to
0^+$ transitions.  Since in the superallowed $0^+ \to 0^+$ transitions
the scalar coupling constant $C_S$ is commensurable with zero we
propose the solution $\bar{C}_S = - C_S = 0$.  For real scalar and
tensor coupling constants and for $\bar{C}_S = - C_S = 0$ and $C_T = -
\bar{C}_T$ we solve Eq.(\ref{eq:6}) and obtain the solution
Eq.(\ref{eq:10}). In the linear approximation we get $C_T = -
\bar{C}_T = 1.11\times 10^{-2}$ and the Fierz interference term equal
to $b = - 1.44 \times 10^{-2}$. We define the correlation coefficients
in terms of the coupling constant $C_T = 1.11\times 10^{-2}$ and the
Fierz interference term $b = -1.44 \times 10^{-2}$. We show that the
contributions of quadratic terms $C^2_T$ are of the standard order
$10^{-4}$. In turn, the contributions of linear terms are of order
$10^{-2} - 10^{-3}$. In section \ref{sec:asymmetry} we analyse i) the
contributions of the Fierz interference term to the electron and
antineutrino asymmetries, defined by the neutron spin and electron and
antineutrino 3--momentum correlations, respectively, and to the
asymmetry, caused by the correlations of the electron and antineutrino
3--momenta, and ii) the averaged values of the correlation
coefficients $N(E_e)$ and $R(E_e)$. The correlation coefficient
$N(E_e)$ defines the neutron--electron spin--spin $\vec{\xi}_n\cdot
\vec{\xi}_e$ correlations, whereas the correlation coefficient
$R(E_e)$ is caused by P--odd (parity odd) and T--odd (time reversal
odd) correlations defined by $\vec{\xi}_n\cdot (\vec{k}_e \times
\vec{\xi}_e)$, where $\vec{\xi}_n$ and $\vec{\xi}_e$ are unit vectors
of the neutron and electron polarizations, and $\vec{k}_e$ is the
electron 3--momentum. We show that the contribution of the Fierz
interference term $b = -1.44 \times 10^{-2}$ does not contradict the
experimental data on the measurements of the correlation coefficients
$A_0$, $B_0$ and also $a_0$, defined to leading order in the large
nucleon mass $M$ expansion \cite{Abele2008,Nico2009}. We would like to
emphasize that the Fierz interference term $b = -1.44 \times 10^{-2}$
does not contradict the result $b = - 0.0028(26)$ extracted by Hardy
and Towner \cite{Hardy2015} from the superallowed $0^+ \to 0^+$
transitions. Indeed, in the superallowed $0^+ \to 0^+$ transitions the
Fierz interference term is defined only by the scalar coupling
constants. At $C_S = - \bar{C}_S$ it is equal to $b \simeq 2C_S$. In
turn, in the neutron $\beta^-$--decay at $C_S = - \bar{C}_S = 0$ the
Fierz interference term is induced by the tensor coupling constants
only. For $C_T = - \bar{C}_T$ we get $b \simeq 6\lambda C_T/(1 + 3
\lambda^2)$ and $b = - 1.44\times 10^{-2}$ for $C_T = 1.11 \times
10^{-2}$ and $\lambda = - 1.2750$ \cite{Abele2008} (see also
\cite{Ivanov2013,Ivanov2017b,Ivanov2018a}). We show that the averaged
value of the correlation coefficient $\langle N(E_e)\rangle$, obtained
in this paper, agrees with the experimental value within one standard
deviations. In turn, the averaged value of the correlation coefficient
$\langle R(E_e)\rangle$ acquires the relative contributions of order
$10^{-4}$, caused by interactions beyond the SM. In section
\ref{sec:schluss} we discuss the obtained results. We argue that the
obtained agreement between theoretical values of the correlation
coefficients, defined by the contributions of the SM to order
$10^{-3}$, the Fierz interference term $b = - 1.44\times 10^{-2}$ and
other linear and quadratic coupling contributions of tensor
interactions beyond the SM, implies an allowance for the neutron to
have dark matter decay modes $n \to \chi + anything$. We would like to
emphasize that the results, obtained in this paper, are
model--independent in the sense that they should be valid and actual
for any model of the solution of the neutron lifetime problem in terms
of neutron dark matter decay modes (see for example
\cite{Barducci2018}).

\section{Electron--energy and angular distribution of neutron 
$\beta^-$--decay with polarized neutron, polarized electron and
  unpolarized proton}
\label{sec:distribution}

The electron--energy and angular distribution of the neutron
$\beta^-$--decay with polarized neutron and electron and unpolarized
proton takes the form \cite{Ivanov2013,Ivanov2017b,Ivanov2018a}
\begin{eqnarray}\label{eq:1}
\hspace{-0.3in}&&\frac{d^5 \lambda_n(E_e,\vec{k}_e,
  \vec{k}_{\nu},\vec{\xi}_n,\vec{\xi}_e)}{dE_e d\Omega_ed\Omega_{\nu}}
=(1 + 3 \lambda^2)\,\frac{G^2_F|V_{ud}|^2}{32\pi^5} \,(E_0 - E_e)^2
\sqrt{E^2_e - m^2_e}\, E_e\,F(E_e, Z = 1)\,\zeta(E_e)\nonumber\\
\hspace{-0.3in}&&\times\,\Big\{1 + b\,\frac{m_e}{E_e} +
a(E_e)\,\frac{\vec{k}_e\cdot \vec{k}_{\nu}}{E_e E_{\nu}} +
A(E_e)\,\frac{\vec{\xi}_n\cdot \vec{k}_e}{E_e} + B(E_e)\,
\frac{\vec{\xi}_n\cdot \vec{k}_{\nu}}{E_{\nu}} +
K_n(E_e)\,\frac{(\vec{\xi}_n\cdot \vec{k}_e)(\vec{k}_e\cdot
  \vec{k}_{\nu})}{E^2_e E}\nonumber\\
\hspace{-0.3in}&&+ Q_n(E_e)\,\frac{(\vec{\xi}_n\cdot
  \vec{k}_{\nu})(\vec{k}_e\cdot \vec{k}_{\nu})}{ E_e E^2_{\nu}} +
D(E_e)\,\frac{\vec{\xi}_n\cdot (\vec{k}_e\times \vec{k}_{\nu})}{E_e
  E_{\nu}} - 3\,\frac{E_e}{M}\,\frac{1 - \lambda^2}{1 + 3
  \lambda^2}\,\Big(\frac{(\vec{k}_e\cdot \vec{k}_{\nu})^2}{E^2_e
  E^2_{\nu}} - \frac{1}{3}\,\frac{k^2_e}{E^2_e}\,\Big)\nonumber\\
\hspace{-0.3in}&& + G(E_e)\,\frac{\vec{\xi}_e \cdot \vec{k}_e}{E_e} +
N(E_e)\,\vec{\xi}_n\cdot \vec{\xi}_e +
Q_e(E_e)\,\frac{(\vec{\xi}_n\cdot \vec{k}_e)( \vec{k}_e\cdot
  \vec{\xi}_e)}{E_e (E_e + m_e)} +
R(E_e)\,\frac{\vec{\xi}_n\cdot(\vec{k}_e \times \vec{\xi}_e)}{E_e} +
\ldots\Big\},
\end{eqnarray}
where $G_F = 1.1664\times 10^{-11}\,{\rm MeV}^{-2}$ and $V_{ud} =
0.97420(21)$ are the Fermi weak coupling constant and the
Cabibbo-Kobayashi--Maskawa (CKM) matrix element \cite{PDG2018},
respectively, $\lambda = - 1.2750(9)$ is a real axial coupling
constant \cite{Abele2008}, $E_0 = (m^2_n - m^2_p + m^2_e)/2 m_n =
1.2927\,{\rm MeV}$ is the end--point energy of the electron spectrum,
calculated for $m_n = 939.5654\,{\rm MeV}$, $m_p = 938.2720\,{\rm
  MeV}$ and $m_e = 0.5110\,{\rm MeV}$ \cite{PDG2018}, $\vec{\xi}_n$
and $\vec{\xi}_e$ are unit polarization vectors of the neutron and
electron, respectively, $F(E_e, Z = 1)$ is the relativistic Fermi
function \cite{Jackson1957a,Blatt1952,Jackson1958,Konopinski1966}
\begin{eqnarray}\label{eq:2}
\hspace{-0.3in}F(E_e, Z = 1 ) = \Big(1 +
\frac{1}{2}\gamma\Big)\,\frac{4(2 r_pm_e\beta)^{2\gamma}}{\Gamma^2(3 +
  2\gamma)}\,\frac{\displaystyle e^{\,\pi
 \alpha/\beta}}{(1 - \beta^2)^{\gamma}}\,\Big|\Gamma\Big(1 + \gamma +
 i\,\frac{\alpha }{\beta}\Big)\Big|^2,
\end{eqnarray}
where $\beta = k_e/E_e = \sqrt{E^2_e - m^2_e}/E_e$ is the electron
velocity, $\gamma = \sqrt{1 - \alpha^2} - 1$, $r_p$ is the electric
radius of the proton.  In the numerical calculations we use $r_p =
0.841\,{\rm fm}$ \cite{Pohl2010}. The Fermi function $F(E_e, Z = 1)$
describes final--state Coulomb proton--electron interaction. Then, $b
$ is the Fierz interference term \cite{Fierz1937}--\cite{Ivanov2018a}
(see also \cite{Gudkov2006} and \cite{Ivanov2013}). The infinitesimal
solid angles $d\Omega_e = \sin\vartheta_ed\vartheta_ed\varphi_e$ and
$d\Omega_{\nu} = \sin\vartheta_{\nu}d\vartheta_{\nu}d\varphi_{\nu}$
are defined relative to the 3--momenta $\vec{k}_e$ and $\vec{k}_{\nu}$
of the decay electron and antineutrino, respectively.

The correlation coefficients $\zeta(E_e)$, $a(E_e)$ and so on, taking
into account the contributions of the SM and scalar and tensor
interactions beyond the SM \cite{Ivanov2013,Ivanov2018a}, and the
Fierz interference term $b$ are given by
\begin{eqnarray}\label{eq:3}
\hspace{-0.3in}\zeta(E_e) = \zeta^{(\rm SM)}(E_e)\big(1 + \zeta^{(\rm
  BSM)}(E_e)\big)\quad,\quad b = \frac{b_F}{\displaystyle 1 +
  \zeta^{(\rm BSM)}(E_e)}\quad,\quad X(E_e) =  \frac{X^{(\rm
    SM)}(E_e) + X^{(\rm BSM)}(E_e)}{\displaystyle 1 + \zeta^{(\rm
    BSM)}(E_e)},
\end{eqnarray}
where $X = a, A, B, D, G, N, Q_e$ and $R$, respectively. The
correlation coefficients $\zeta^{(\rm SM)}(E_e)$ and $X^{(\rm
  SM)}(E_e)$ for $X = a, A$ and so on are calculated within the SM
including the complete set of corrections caused by the weak magnetism
and proton recoil of order $O(E_e/M)$ and radiative corrections of
order $O(\alpha/\pi)$ \cite{Ivanov2013,Ivanov2017b}. In turn, the
correlation coefficients $b_F$, $\zeta^{(\rm BSM)}(E_e)$ and $X^{(\rm
  BSM)} = a, A$ and so on are defined by \cite{Ivanov2013,Ivanov2018a}
\begin{eqnarray*}
b_F &=& \frac{1}{1 + 3\lambda^2}\,{\rm Re}((C_S - \bar{C}_S) + 3
\lambda\,(C_T -\bar{C}_T)),\nonumber\\
\hspace{-0.3in}\zeta^{(\rm BSM)}(E_e) &=& \frac{1}{2}\,\frac{1}{1 +
  3\lambda^2}\,( |C_S|^2 + |\bar{C}_S|^2 + 3 |C_T|^2 +
3|\bar{C}_T|^2),\nonumber\\ a^{(\rm BSM)}(E_e) &=&-
\frac{1}{2}\,\frac{1}{1 + 3 \lambda^2}\,\big(|C_S|^2 + |\bar{C}_S|^2 -
|C_T|^2 + |\bar{C}_T|^2 \big),\nonumber\\ A^{(\rm BSM)}(E_e) &=&-
\frac{1}{1 + 3 \lambda^2}\,{\rm Re}\big( C_S \bar{C}^*_T + \bar{C}_S
C^*_T + 2 C_T \bar{C}^*_T\big),\nonumber\\ B^{(\rm BSM)}(E_e) &=&-
\frac{1}{1 + 3 \lambda^2}\,{\rm Re}\big(2 C_T\bar{C}^*_T -
C_S\bar{C}^*_T - \bar{C}_S C^*_T\big) -
b_N\,\frac{m_e}{E_e},\nonumber\\ 
D^{(\rm BSM)}(E_e) &=& \frac{1}{1 + 3
  \lambda^2}\,{\rm Im}\big( C_S C^*_T + \bar{C}_S
\bar{C}^*_T\big),\nonumber\\ 
G^{(\rm BSM)}(E_e) &=& - \frac{1}{1 +
  3\lambda^2}\,{\rm Re}(C_S\bar{C}^*_S + 3
C_T\bar{C}^*_T),\nonumber\\ 
N^{(\rm BSM)}(E_e)&=& \frac{m_e}{E_e}\,
\frac{1}{1 + 3\lambda^2}\,{\rm Re}(C_S C^*_T + \bar{C}_S \bar{C}^*_T +
|C_T|^2 + |\bar{C}_T|^2) + b_E,\nonumber\\
 \end{eqnarray*}
\begin{eqnarray}\label{eq:4}
Q^{(\rm BSM)}_e(E_e)&=&
\frac{1}{1 + 3\lambda^2}\,{\rm Re}(C_S C^*_T + \bar{C}_S \bar{C}^*_T +
|C_T|^2 + |\bar{C}_T|^2) - b_E, \nonumber\\
 R^{(\rm BSM)}(E_e)&=&
\frac{1}{1 + 3\lambda^2}\,{\rm Im}(\lambda (C_S - \bar{C}_S) + (1 + 2
\lambda)(C_T - \bar{C}_T)),\nonumber\\ b_E &=& \frac{1}{1 +
  3\lambda^2}\,{\rm Re}(\lambda\, (C_S - \bar{C}_S) + (1 + 2\lambda)\,
(C_T -\bar{C}_T)),\nonumber\\ b_N &=& \frac{1}{1 + 3\lambda^2}\,{\rm
  Re}(\lambda\, (C_S - \bar{C}_S) + (1 - 2\lambda)\, (C_T
-\bar{C}_T)).
\end{eqnarray}
Now we may proceed to the analysis of contributions of scalar and
tensor interactions beyond the SM to the rate of the neutron decay
modes $n \to p + anything$, measured in beam experiments, and
correlation coefficients under consideration.

\section{Neutron lifetime measured in beam experiments}
\label{sec:lifetime}

Using the results, obtained in \cite{Ivanov2013,Ivanov2018a}, the
neutron lifetime is given by the following expression

\begin{eqnarray}\label{eq:5}
\hspace{-0.3in} \frac{1}{\tau_n} &=& (1 +
3\lambda^2)\,\frac{G^2_F|V_{ud}|^2}{2\pi^3}f_n \Big(1 +
\frac{1}{2}\,\frac{1}{1 + 3\lambda^2}\,\Big( ({\rm Re}C_S)^2 + ({\rm
  Im}C_S)^2 + ({\rm Re}\bar{C}_S)^2 + ({\rm Im}\bar{C}_S)^2 + 3({\rm
  Re}C_T)^2 \nonumber\\
\hspace{-0.3in}&+& 3({\rm Im}C_T)^2 + 3 ({\rm Re}\bar{C}_T)^2 + 3({\rm
  Im}\bar{C}_T)^2\Big) + b_F\Big\langle
\frac{m_e}{E_e}\Big\rangle_{\rm SM}\Big),
\end{eqnarray}
where $\langle m_e/E_e\rangle_{\rm SM} = 0.6556$
\cite{Ivanov2013,Ivanov2018a}, and $f_n = 0.0616\,{\rm MeV^5}$ is the
Fermi integral \cite{Ivanov2013,Ivanov2018a}. Since $(1 +
3\lambda^2)G^2_F|V_{ud}|^2f_n/2\pi^3 = 1/879.6 \,{\rm s^{-1}}$
\cite{Ivanov2013}, we get
\begin{eqnarray}\label{eq:6}
\hspace{-0.3in}&&\frac{1}{2}\,\frac{1}{1 + 3\lambda^2}\,\Big( ({\rm
  Re}C_S)^2 + ({\rm Im}C_S)^2 + ({\rm Re}\bar{C}_S)^2 + ({\rm
  Im}\bar{C}_S)^2 + 3({\rm Re}C_T)^2 + 3({\rm Im}C_T)^2 + 3 ({\rm
  Re}\bar{C}_T)^2 + 3({\rm Im}\bar{C}_T)^2\Big) =\nonumber\\ 
\hspace{-0.3in}&&= - b_F\Big\langle \frac{m_e}{E_e}\Big\rangle_{\rm
  SM} - \frac{\Delta \tau_n}{\tau_n},
\end{eqnarray}
where $\Delta \tau_n = 8.4\,{\rm s}$, $\tau_n = 888.0\,{\rm s}$ and
$\Delta \tau_n/\tau_n = 9.46\times 10^{-3}$.  It is seen that the
correlation coefficient $b_F$, defining the sign of the Fierz
interference term $b$ (see Eq.(\ref{eq:3}) should be negative.

\section{Correlation coefficients and numerical analysis of  scalar 
and tensor coupling constants}
\label{sec:coefficient}

The aim of this paper is to find anyone plausible solution for the
scalar and tensor coupling constants of interactions beyond the SM,
which should be compatible with present time accuracy of the
definition of the correlation coefficients of the neutron
$\beta^-$--decay. As a first step of the analysis of the contributions
of scalar and tensor interaction beyond the SM we follow
\cite{Severijns2006} and \cite{Ivanov2018a} and set i) zero the
imaginary parts of the scalar and tensor coupling constants and ii)
$\bar{C}_j = - C_j$ for $ j = S, T$. This gives
\begin{eqnarray}\label{eq:7}
\hspace{-0.3in}&&\frac{1}{1 + 3\lambda^2}\,\big(C^2_S + 3 C^2_T\big) =
- b_F\Big\langle \frac{m_e}{E_e}\Big\rangle_{\rm SM} - \frac{\Delta
  \tau_n}{\tau_n}
\end{eqnarray}
and 
\begin{eqnarray}\label{eq:8}
\hspace{-0.3in}b_F = \frac{2}{1 + 3\lambda^2}\,\big(C_S + 3
\lambda\,C_T \big) \, &,&\, \zeta^{(\rm BSM)}(E_e) = \frac{1}{1 +
  3\lambda^2}\,\big( C^2_S + 3 C^2_T\big)\nonumber\\ a^{(\rm
  BSM)}(E_e) = - \frac{1}{1 + 3 \lambda^2}\,\big(C^2_S - C^2_T\big)
\, &,&\, A^{(\rm BSM)}(E_e) = \frac{2}{1 + 3 \lambda^2}\,\big( C_S
C_T + C^2_T\big),\nonumber\\ B^{(\rm BSM)}(E_e) = \frac{2}{1 + 3
  \lambda^2}\,\big(C^2_T - C_S C_T\big) - b_N\,\frac{m_e}{E_e}
\,&,&\, D^{(\rm BSM)}(E_e) = 0,\nonumber\\ 
\hspace{-0.3in}G^{(\rm
  BSM)}(E_e) = \frac{1}{1 + 3\lambda^2}\,\big(C^2_S + 3 C^2_T \big)
\, &,&\, N^{(\rm BSM)}(E_e) = \frac{m_e}{E_e}\, \frac{2}{1 +
  3\lambda^2}\,\big(C_S C_T + C^2_T\big) +
b_E,\nonumber\\ 
\hspace{-0.3in}Q^{(\rm BSM)}_e(E_e) = \frac{2}{1 +
  3\lambda^2}\, \big(C_S C_T + C^2_T\big) - b_E \, &,&\, R^{(\rm
  BSM)}(E_e) = 0,\nonumber\\ 
\hspace{-0.3in}b_E = \frac{2}{1 +
  3\lambda^2}\,\big(\lambda\,C_S + (1 +
2\lambda)\,C_T\big) \, &,&\, b_N  = \frac{2}{1 +
  3\lambda^2}\,\big(\lambda\,C_S + (1 - 2\lambda)\, C_T \big).
\end{eqnarray}
The simplest solution, which sticks a mile and is comparable with the
constraints $|C_S| = 0.0014(12)$ and $|C_S| = 0.0014(13)$ obtained by
Hardy and Towner \cite{Hardy2015} and Gonz\'alez-Alonso {\it et al.}
\cite{Severijns2018} from the superallowed $0^+ \to 0^+$ transitions,
is $C_S = - \bar{C}_S = 0$. Setting $C_S = 0$ we transcribe the
correlation coefficients into the form
\begin{eqnarray}\label{eq:9}
\hspace{-0.15in}a(E_e) = \frac{\displaystyle a^{(\rm SM)}(E_e) +
  \frac{1}{1 + 3\lambda^2}\,C^2_T}{\displaystyle 1 + \frac{3}{1 +
    3\lambda^2}\,C^2_T}\, &,&\, A(E_e) = \frac{\displaystyle A^{(\rm
    SM)}(E_e) + \frac{2}{1 + 3\lambda^2}\,C^2_T}{\displaystyle 1 +
  \frac{3}{1 + 3\lambda^2}\,C^2_T },\nonumber\\
\hspace{-0.15in}B(E_e) = \frac{\displaystyle B^{(\rm SM)}(E_e) +
  \frac{2}{1 + 3 \lambda^2}\,C^2_T -
  b_N\,\frac{m_e}{E_e}}{\displaystyle 1 + \frac{3}{1 +
    3\lambda^2}\,C^2_T }\, &,&\, D(E_e) = \frac{D^{(\rm
    SM)}(E_e)}{\displaystyle 1 + \frac{3}{1 + 3\lambda^2}\, C^2_T
},\nonumber\\
\hspace{-0.15in} G(E_e) = \frac{\displaystyle G^{(\rm SM)}(E_e) +
  \frac{3}{1 + 3\lambda^2}\,C^2_T }{\displaystyle 1 + \frac{3}{1 +
    3\lambda^2}\, C^2_T}\, &,&\, N(E_e) = \frac{\displaystyle N^{(\rm
    SM)}(E_e) + \frac{2}{1 + 3\lambda^2}\,C^2_T\,\frac{m_e}{E_e} + b_E
}{\displaystyle 1 + \frac{3}{1 + 3\lambda^2}\,C^2_T },\nonumber\\
\hspace{-0.15in} Q_e(E_e) = \frac{\displaystyle Q^{(\rm SM)}_e(E_e) +
  \frac{2}{1 + 3\lambda^2}\,C^2_T - b_E}{\displaystyle 1 + \frac{3}{1
    + 3\lambda^2}\,C^2_T }\, &,&\, R(E_e) = \frac{R^{(\rm
    SM)}(E_e)}{\displaystyle 1 + \frac{3}{1 +
    3\lambda^2}\,C^2_T},\nonumber\\ b_F = \frac{ 6\lambda}{~1 +
  3\lambda^2}\,C_T\;,\; b_N = 2\,\frac{1 - 2\lambda}{~~1 +
  3\lambda^2}\,C_T\,&,&\, b_E = 2\,\frac{1 + 2\lambda}{~~1 +
  3\lambda^2}\,C_T.
\end{eqnarray}
At $C_S = 0$ Eq.(\ref{eq:7}) reduces to the quadratic algebraical
equation with the solution
\begin{eqnarray}\label{eq:10}
C_T = - \lambda \Big\langle \frac{m_e}{E_e}\Big\rangle_{\rm SM}\Bigg(1
- \sqrt{1 - \frac{1 + 3\lambda^2}{3\lambda^2}\,\frac{\Delta
    \tau_n}{\tau_n}\,\Big\langle \frac{m_e}{E_e}\Big\rangle^{-2}_{\rm
    SM}}\,\Bigg),
\end{eqnarray}
where we have chosen only the solution obeying the constraint $|C_T|
\ll 1$. In the linear approximation we get
\begin{eqnarray}\label{eq:11}
C_T &=&- \frac{1 + 3\lambda^2}{6 \lambda}\,\frac{\Delta
  \tau_n}{\tau_n}\,\Big\langle \frac{m_e}{E_e}\Big\rangle^{-1}_{\rm
  SM} = 1.11 \times 10^{-2}\;,\, \frac{3}{1 + 3\lambda^2}\,C^2_T =
6.27\times 10^{-5},\nonumber\\ 
b_F &=& - \frac{\Delta
  \tau_n}{\tau_n}\,\Big\langle \frac{m_e}{E_e}\Big\rangle^{-1}_{\rm
  SM} = - 1.44\times 10^{-2}\;,\; b_N = - \frac{1 -
  2\lambda}{3\lambda}\,\frac{\Delta \tau_n}{\tau_n}\,\Big\langle
\frac{m_e}{E_e}\Big\rangle^{-1}_{\rm SM} = 1.34\times 10^{-2},
\nonumber\\ 
b_E &=& - \frac{1 + 2\lambda}{3\lambda}\,\Big\langle
\frac{m_e}{E_e}\Big\rangle^{-1}_{\rm SM} = - 5.85\times 10^{-3}.
\end{eqnarray}
Plugging Eq.(\ref{eq:11}) into Eq.(\ref{eq:9}) we obtain the
following correlation coefficients, corrected by the contributions of
tensor interactions beyond the SM:
\begin{eqnarray}\label{eq:12}
\hspace{-0.15in}a(E_e) &=& a^{(\rm SM)}(E_e)\big(1 - 6.27\times
10^{-5}\big) + 2.09\times 10^{-5},\nonumber\\
 A(E_e) &=& A^{(\rm
  SM)}(E_e)\,\big(1 - 6.27\times 10^{-5}\big) + 4.18\times
10^{-5},\nonumber\\
\hspace{-0.15in}B(E_e) &=& B^{(\rm SM)}(E_e) \Big(1 - 1.34 \times
10^{-2}\,\frac{m_e}{E_e}\Big)- 2.09\times 10^{-5},\nonumber\\ 
D(E_e)
&=& D^{(\rm SM)}(E_e)\big(1 - 6.27\times
10^{-5}\big),\nonumber\\ 
\hspace{-0.15in} G(E_e) &=& G^{(\rm
  SM)}(E_e)\,\big(1 + 1.25\times 10^{-4}\big),\nonumber\\ 
N(E_e) &=&
N^{(\rm SM)}(E_e)\,\big(1 - 6.27\times 10^{-5}\big) + 4.18\times
10^{-5}\,\frac{m_e}{E_e} - 5.85\times
10^{-3},\nonumber\\ \hspace{-0.15in} Q_e(E_e) &=& Q^{(\rm
  SM)}_e(E_e)\,\big(1 - 6.27\times 10^{-5}\big) + 5.89\times
10^{-3},\nonumber\\ R(E_e) &=& R^{(\rm SM)}\,\big(1 - 6.27\times
10^{-5}\big).
\end{eqnarray}
Thus, we have shown that there exists at least one solution $C_S = -
\bar{C}_S = 0$ and $C_T = - \bar{C}_T = 1.11\times 10^{-2}$ for real
scalar and tensor coupling constants with the Fierz interference term
$b = - 1.44 \times 10^{-2}$, which does not contradict the constraints
$|C_S| = 0.0014(13)$ \cite{Hardy2015} and $|C_S| = 0.0014(12)$
\cite{Severijns2018} obtained from the superallowed $0^+ \to 0^+$
transitions and determines reasonable contributions of interactions
beyond the SM to the correlation coefficients $a(E_e)$ and $A(E_e)$ of
order $10^{-4}$ \cite{Abele2016} and to the correlation coefficients
$G(E_e)$, $N(E_e)$ and $Q_e(E_e)$ of order $10^{-3}$. In turn, the
correlation coefficient $B(E_e) \sim 1$ acquires the correction of
order $10^{-2}$.

Now we have to compare the obtained results with the experimental
data.  For this aim we have to analyse the asymmetries of the neutron
$\beta^-$--decay and the averaged values of correlation coefficients.

\section{Asymmetries and averaged values of correlation coefficients}
\label{sec:asymmetry}

\subsection{Electron asymmetry of neutron $\beta^-$--decay}

\begin{figure}
\includegraphics[height=0.20\textheight]{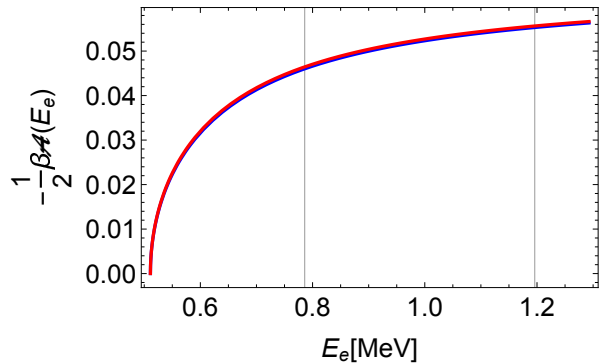}
  \caption{The theoretical electron asymmetry, calculated in the SM
    (blue curve) at $b = C_T = 0$ and with interactions beyond the SM
    (red curve) for $C_T = 1.11 \times 10^{-2}$ and $b = - 1.44\times
    10^{-2}$, respectively. }
\label{fig:fig1}
\end{figure}

The most sensitive asymmetry of the neutron $\beta^-$--decay is the
electron asymmetry, caused by correlations of the neutron spin
$\vec{\xi}_n$ and the electron 3--momentum $\vec{k}_e$ and described
by the scalar product $\vec{\xi}_n\cdot \vec{k}_e$. The experimental
electron asymmetry $A_{\exp}(E_e)$ of electrons emitted forward and
backward with respect to the neutron spin $\vec{\xi}_e$ into the solid
angle $\Delta \Omega_{12} = 2\pi (\cos\theta_1 - \cos\theta_2)$ with
$0 \le \varphi \le 2\pi$ and $\theta_1 \le \theta_e \le \theta_2$ is
equal to \cite{Ivanov2013,Ivanov2018a}
\begin{eqnarray}\label{eq:13}
A_{\exp}(E_e) = \frac{1}{2}\,\beta\,\bar{A}_W(E_e) P_n (\cos\theta_1 +
\cos\theta_2),
\end{eqnarray}
where $P_n = |\vec{\xi}_n| \le 1$ is the neutron spin polarization,
and the correlation coefficient $\bar{A}_W(E_e)$ is
\cite{Ivanov2013,Ivanov2018a}
\begin{eqnarray}\label{eq:14}
\bar{A}_W(E_e) = \frac{\displaystyle {\cal A}_W(E_e) }{\displaystyle 1
  + b\,\frac{m_e}{E_e}}.
\end{eqnarray}
The correlation coefficient ${\cal A}_W(E_e)$ can be obtained from the
correlation coefficient $A(E_e)$ given by Eq.(\ref{eq:9}) and
Eq.(\ref{eq:12}), respectively, with $A^{(\rm SM)}(E_e)$ replaced by
$A^{(\rm SM)}_W(E_e)$ taken in the form
\begin{eqnarray}\label{eq:15}
A^{(\rm SM)}_W(E_e) = \Big(1 +
\frac{\alpha}{\pi}\,f_n(E_e)\Big)\,A_W(E_e),
\end{eqnarray}
where the function $f_n(E_e)$ defines the radiative corrections,
calculated by Shann \cite{Shann1971}, and the function $A_W(E_e)$ has
been calculated by Bilen'kii {\it et al.} \cite{Bilenky1959} and by
Wilkinson \cite{Wilkinson1982} by taking into account the
contributions of order $O(E_e/M)$ caused by the weak magnetism and
proton recoil (see also \cite{Ivanov2013}). Following \cite{Abele2013}
we plot in Fig.\,\ref{fig:fig1} the function $- \frac{1}{2} \beta
{\cal A}(E_e)$ in the electron energy region $m_e \le E_e \le E_0$,
where the electron asymmetry calculated in the SM at $b = C_T = 0$ is
given by the blue curve, whereas the electron asymmetry calculated
with the account for the contributions of interactions beyond the SM
at $C_T = 1.11\times 10^{-2}$ and $b = 6\lambda C_T/(1 + 3\lambda^2) =
- 1.44\times 10^{-2}$ is presented by the red curve. One may see that
these two theoretical curves cannot be practically distinguished in
experiments.
The results presented by the theoretical curves in
Fig.\,\ref{fig:fig1} can be also confirmed by the following estimates.

We propose to estimate the correlation coefficient $\bar{A}_W(E_e)$,
calculated to leading order in the large nucleon mass $M$ expansion
and at the neglect of the radiative corrections. The result is equal
to
\begin{eqnarray}\label{eq:16}
\lim_{M \to \infty}\bar{A}_W(E_e) = \frac{\displaystyle A_0\Big(1 -
  \frac{3}{1 + 3\lambda^2}\,C^2_T\Big) + \frac{2}{1 +
    3\lambda^2}\,C^2_T}{\displaystyle 1 + b \,\frac{m_e}{E_e}}\quad,
\quad A_0 = - 2\,\frac{\lambda (1 + \lambda)}{1 + 3\lambda^2} = -
0.11933,
\end{eqnarray}
where $A_0 = - 0.11933$ is calculated at $\lambda = -1.2750$
\cite{Abele2008}. For $C_T = 1.11\times 10^{-2}$ and $b = 6\lambda
C_T/(1+ 3\lambda^2) = - 1.44 \times 10^{-2}$ we get $- 0.12102 \le
\lim_{M \to \infty}\bar{A}_W(E_e)\le - 0.11997$ for the
electron--energy region $m_e \le E_e \le E_0$. These values agree well
with recent experimental values $A_0 = - 0.12015(34)_{\rm
  stat.}(63)_{\rm syst.}$ and $(A_0)_{\exp} = - 0.12054(44)_{\rm
  stat.}(68)_{\rm syst.}$ \cite{Brown2018}.

The analogous estimate we may make for the correlation coefficient
$\bar{a}(E_e) = a(E_e)/(1 + b\,m_e/E_e)$, describing the asymmetry of
electron--antineutrino 3--momentum correlations, defined by the scalar
product $\vec{k}_e\cdot \vec{k}_{\nu}$. To leading order in the large
nucleon mass $M$ expansion we get
\begin{eqnarray}\label{eq:17}
\lim_{M \to \infty}\bar{a}(E_e) = \frac{\displaystyle a_0\Big(1 -
  \frac{3}{1 + 3\lambda^2}\,C^2_T\Big) + \frac{1}{1 +
    3\lambda^2}\,C^2_T}{\displaystyle 1 + b \frac{m_e}{E_e}}\quad,
\quad a_0 = \frac{1 - \lambda^2}{1 + 3\lambda^2} = - 0.1065,
\end{eqnarray}
where $a_0 = - 0.10645$ is calculated at $\lambda = - 1.2750$. At $C_T
= 1.11\times 10^{-2}$ and $b = 6\lambda C_T/(1+ 3\lambda^2) =
-1.44\times 10^{-2}$ the correlation coefficient $\lim_{M \to
  \infty}\bar{a}(E_e)$ is constrained by $- 0.1080 \le \lim_{M \to
  \infty}a(E_e)\le - 0.1071$ in the electron--energy region $m_e \le
E_e \le E_0$.  The values $- 0.1080 \le \lim_{M \to
  \infty}\bar{a}(E_e)\le - 0.1071$ agree well with recent experimental
value $(a_0)_{\exp} = - 0.1090 \pm 0.0030_{\rm stat.} \pm 0.0028_{\rm
  sys.}$ from the aCORN experiment reported in \cite{Nico2017}.

\subsection{Antineutrino asymmetry of neutron $\beta^-$--decay}

The antineutrino asymmetry of the neutron $\beta^-$--decay is caused
by correlations of the neutron spin $\vec{\xi}_n$ and antineutrino
3--momentum $\vec{k}_{\nu}$, defined by the scalar product
$\vec{\xi}_n \cdot \vec{k}_{\nu}$. Following \cite{Schumann2007} we
define the antineutrino asymmetry $B_{\exp}(E_e)$ as follows
\cite{Ivanov2013}
\begin{eqnarray}\label{eq:18}
  B_{\exp}(E_e) = \frac{N^{--}(E_e) - N^{++}(E_e)}{N^{--}(E_e) +
  N^{++}(E_e)}.
\end{eqnarray}
This expression determines the asymmetry of the emission of the
antineutrinos into the forward and backward hemisphere with respect to
the neutron spin, where $N^{\mp \mp}(E_e)$ is the number of events of
the emission of the electron--proton pairs as functions of the
electron energy $E_e$. The signs $(++)$ and $(--)$ show that the
electron--proton pairs were emitted parallel $(++)$ and antiparallel
$(--)$ to a direction of the neutron spin. This means that
antineutrinos were emitted antiparallel $(++)$ and parallel $(--)$ to
a direction of the neutron spin.  The number of events $N^{--}(E_e)$
and $N^{++}(E_e)$ are defined by the electron--energy and angular
distribution of the neutron $\beta^-$--decay, integrated over the
forward and backward hemisphere relative to the neutron spin,
respectively \cite{Ivanov2013}. The analytical expression for
$B_{\exp}(E_e)$ is given by \cite{Ivanov2013}
\begin{eqnarray}\label{eq:19}
B^{(r \le 1)}_{\exp}(E_e) = \frac{2 P}{3}\,\frac{\displaystyle (3 -
  r^2) \bar{B}(E_e) - (3 - 2 r)\beta \bar{A}(E_e) + \Big(1 -
  \frac{3}{5} r^2\Big)\,\beta^2 \bar{K}_n(E_e)- \Big(1 - \frac{2}{5}
  r^3\Big) \beta \bar{Q}_n(E_e)}{\displaystyle (4 - 2r) - \Big(1 -
  \frac{1}{2}\,r^2\Big) \beta \bar{a}(E_e)-
  \frac{1}{2}\,\bar{a}_0(E_e)\beta^2\, r(1 - r^2)\,\frac{E_e}{M}}
\end{eqnarray}
and
\begin{eqnarray}\label{eq:20}
\hspace{-0.3in}B^{(r\ge 1)}_{\exp}(E_e) =
\frac{2P}{3}\,\frac{\displaystyle \bar{B}(E_e) -
  \frac{1}{2}\,\Big(\bar{A}(E_e) + \frac{3}{5}\bar{Q}_n(E_e)\Big)\,
  \frac{\beta}{r} + \frac{1}{5}
  \bar{K}_n(E_e)\,\frac{\beta^2}{r^2}}{\displaystyle 1 -
  \bar{a}(E_e)\,\frac{\beta}{4 r} +
  \frac{1}{4}\,\bar{a}_0(E_e)\beta^2\,\Big(1 -
  \frac{1}{r^2}\Big)\frac{E_e}{M}},
\end{eqnarray}
where $r = k_e/E_{\nu} = k_e/(E_0 - E_e)$ \cite{Schumann2007} (see
also \cite{Ivanov2013}). For $r \le 1$ and $r \ge 1$ the electron
kinetic energies are restricted by $0 \le T_e \le (E_0 - m_e)^2/2 E_0
= 236\,{\rm keV}$ and $(E_0 - m_e)^2/2 E_0 = 236\,{\rm keV}\le T_e \le
E_0 - m_e$, respectively. At $r = 1$ or at the electron kinetic energy
$T_e = (E_0 - m_e)^2/2 E_0 = 236\,{\rm keV}$ the antineutrino
asymmetry given by Eq.(\ref{eq:19}) and Eq.(\ref{eq:20}) is continues.
The correlation coefficients $\bar{X}(E_e)$ for $X = a, A, B, Q_n$ and
$K_n$ take the form $\bar{X}(E_e) = X(E_e)/(1 + b\,m_e/E_e)$, where
$X(E_e)$ are defined by Eq.(\ref{eq:9}) and Eq.(\ref{eq:12}),
respectively, with $X^{(\rm SM)}(E_e)$ calculated in
\cite{Ivanov2013}. Then, $\bar{a}_0(E_e)$ is given by $\bar{a}_0(E_e)
= \big(a_0(1 - 6.27\times 10^{-5}) + 2.09 \times 10^{-5}\big)/(1 +
b\,m_e/E_e)$ (see Eq.(\ref{eq:12})), where the numerator is calculated
to leading order in the large nucleon mass $M$ expansion. The main
contribution to the antineutrino asymmetry $B_{\exp}(E_e)$ comes from
the term $B^{(\rm SM)}(E_e) \sim 1$, which does not depend on the
tensor coupling constant $C_T$.  Since the correlation coefficients
$a^{(\rm SM)}(E_e)$ and $A^{(\rm SM)}(E_e)$ are of order $a^{(\rm
  SM)}(E_e) \sim A^{(\rm SM)}(E_e) \sim - 0.1$ and the values of
correlations coefficients $K^{(\rm SM)}_n(E_e)$ and $Q^{(\rm
  SM)}_n(E_e)$ are of order $10^{-3}$, the contribution of the Fierz
interference term $b = - 1.44 \times 10^{-2}$ to the antineutrino
asymmetry is of about $0.1\,\%$ or even smaller. The experimental
value of the correlation coefficients $(B_0)_{\exp} = 0.9802(50)$
\cite{Schumann2007} is defined with an accuracy of about
$0.5\,\%$. This means that at the present level of experimental
accuracy the antineutrino asymmetry of the neutron $\beta^-$--decay is
not sensitive to the contribution of the tensor interactions beyond
the SM with $C_T = 1.11\times 10^{-2}$ and the Fierz interference term
$b = 6 \lambda C_T/(1 + 3\lambda^2) = - 1.44 \times 10^{-2}$.

\subsection{Averaged values of correlation coefficients}

In the neutron $\beta^-$--decay with polarized neutron and electron
and unpolarized proton the averaged values have been measured only for
the correlation coefficients $N(E_e)$ and $R(E_e)$: $N_{\exp} =
\langle N(E_e)\rangle = 0.067\pm 0.011\pm 0.004$ and $R_{\exp} =
\langle R(E_e)\rangle = 0.004 \pm 0.012 \pm 0.005$
\cite{Kozela2012}. Using Eq.(\ref{eq:12}) and the results, obtained in
\cite{Ivanov2017b,Ivanov2018a}, we get
\begin{eqnarray}\label{eq:21}
\hspace{-0.3in}\langle N(E_e)\rangle = 0.07185\quad,\quad \langle
R(E_e)\rangle = 0.00089.
\end{eqnarray}
The averaged value of the correlation coefficient $\langle
N(E_e)\rangle = 0.07185$ agrees with the experimental one within one
standard deviation. In turn, as it follows from Eq.(\ref{eq:12}) the
averaged value of the correlation coefficient $\langle R(E_e)\rangle$
can acquire only the relative correction of order $10^{-4}$, caused by
tensor interactions beyond the SM. This does not contradict the
experimental data by \cite{Kozela2012}.

\section{Discussion}
\label{sec:schluss}

The main aim of this paper is to show that the fit of the rate of the
neutron decay modes $n \to p + anything$, measured in beam
experiments, by the Fierz interference term $b = - 1.44\times 10^{-2}$
does not contradict contemporary experimental data on the values of
the correlation coefficients of the neutron $\beta^-$--decay with
polarized neutron, polarized electron and unpolarized proton. We have
found that there exist as minimum one solution of our interest for
real scalar and tensor coupling constants $C_S = - \bar{C}_S = 0$ and
$C_T = -\bar{C}_T = 1.11 \times 10^{-2}$. This solution agrees well
with the constraint $|C_S| = 0.0014(13)$ and $|C_S| = 0.0014(12)$
extracted from the superallowed $0^+ \to 0^+$ transitions by Hardy and
Towner \cite{Hardy2015} and Gonz\'alez-Alonso {\it et al.}
\cite{Severijns2018}, respectively, and defines the Fierz interference
term $b = - 1.44\times 10^{-2}$. The contributions of this solution of
order $10^{-4} - 10^{-2}$ to the correlation coefficients of the
electron--energy and angular distributions of the neutron
$\beta^-$--decays do not contradict contemporary experimental data on
the correlation coefficients and asymmetries of the neutron
$\beta^-$--decays with polarized neutron, polarized electron and
unpolarized proton. We would like to emphasize that the problem of
such a fit is not related only to our model \cite{Ivanov2018d} for the
solution of the neutron lifetime anomaly, but it appears always as
soon as the neutron dark matter decay modes $n \to \chi + anything$
would have been observed. For example, the neutron dark matter decay
mode $n \to \chi + e^- + e^+$, proposed by Fornal and Grinstein
\cite{Fornal2018}, can be in principle observed in the region
$100\,{\rm keV} < T_{-+}$ of kinetic energies by the UCNA and PERKEO
II Collaborations \cite{Sun2018,Ivanov2018d}. This concerns also the
decay mode $n \to \chi + \phi$ \cite{Fornal2018,Fornal2018a}, which is
not still investigated experimentally. Hence, one may argue that in
any model of neutron dark matter decays (see, for example,
\cite{Barducci2018}) our analysis of the rate of the neutron decay
modes $n \to p + anything$, measured in beam experiments, and the
correlation coefficients should be meaningful and actual.

The results of our analysis on the rate of the neutron decay modes $n
\to p + anything$, measured in beam experiments, and correlation
coefficients of electron--energy and angular distributions of the
neutron $\beta^-$--decays can be interpreted as an allowance for the
neutron to have the dark matter decay modes $n \to \chi + anything$,
which have been analysed in \cite{Ivanov2018d} in the physical phase
of a quantum field theory model with $SU_L(2) \times U_Y(1) \times
U'_R(1) \times U''_L(1)$ gauge symmetry or in any other model of
neutron dark matter decays (see, for example, \cite{Barducci2018}). In
our model the traces of the dark matter fermions $\chi$ with mass
$m_{\chi} < m_n$ can be searched in terrestrial laboratories through
the measurements of the differential cross section for the low--energy
inelastic electron--neutron scattering $e^- + n \to \chi + e^-$
\cite{Ivanov2018d} and the triple--differential cross section for the
electrodisintegration of the deuteron into dark matter fermions and
protons $e^- + d \to \chi + p + e^-$ close to threshold
\cite{Ivanov2018e}.

Our analysis of the rate of the neutron decay modes $n \to p + {\rm
  anything}$, measured in beam experiments, and the correlation
coefficients of the neutron $\beta^-$--decay, taking into account the
complete set of corrections of order $10^{-3}$, caused by the weak
magnetism and proton recoil of order $O(E_e/M)$ and radiative
corrections of order $O(\alpha/\pi)$, and the tensor coupling
constants $C_T = - \bar{C}_T = 1.11\times 10^{-2}$ and the Fierz
interference term $b = -1.44\times 10^{-2}$, does diminish an
important role of the SM corrections of order $10^{-5}$, which has
been pointed out in
\cite{Ivanov2017b,Ivanov2018a,Ivanov2017c,Ivanov2018b}. As has been
shown in \cite{Ivanov2017b} the SM corrections of order $10^{-5}$
concern i) Wilkinson's corrections \cite{Wilkinson1982}, i.e. the
higher order corrections caused by 1) the proton recoil in the Coulomb
electron--proton final--state interaction, 2) the finite proton
radius, 3) the proton--lepton convolution and 4) the higher--order
{\it outer} radiative corrections, and then ii) the higher order
corrections defined by 1) the radiative corrections of order
$O(\alpha^2/\pi^2)$, calculated to leading order in the large nucleon
mass expansion, 2) the radiative corrections of order $O(\alpha
E_e/M)$, calculated to next--to--leading order in the large nucleon
mass expansion, which depend strongly on contributions of hadronic
structure of the nucleon \cite{Ivanov2017c,Ivanov2018b}, and 3) the
corrections of the weak magnetism and proton recoil of order
$O(E^2_e/M^2)$, calculated to next--to--next--to--leading order in the
large nucleon mass expansion \cite{Ivanov2018a,Ivanov2017c}. These
theoretical corrections should provide for the analysis of
experimental data of "discovery" experiments the required $5\sigma$
level of experimental uncertainties of a few parts in $10^{-5}$
\cite{Ivanov2017b}.

An important role of strong low--energy interactions and contributions
of hadronic structure of the nucleon for a correct gauge invariant
calculation of radiative corrections of order $O(\alpha E_e/M)$ and
$O(\alpha^2/\pi^2)$ as functions of the electron energy $E_e$ has been
pointed out in \cite{Ivanov2017c,Ivanov2018a}. This agrees well with
Weinberg's assertion about important role of strong low--energy
interactions in decay processes \cite{Weinberg1957}. A procedure for
the calculation of these radiative corrections to the neutron
$\beta^-$--decays with a consistent account for contributions of
strong low--energy interactions, leading to gauge invariant observable
expressions dependent on the electron energy $E_e$ determined at the
confidence level of Sirlin's radiative corrections \cite{Sirlin1967},
has been proposed in \cite{Ivanov2017c,Ivanov2018b}.

The calculation of the SM corrections of order $10^{-5}$ should also
give a theoretical background for the experimental analysis of the
corrections caused by the second class currents \cite{Weinberg1958},
which has been analysed in the neutron $\beta^-$--decay with polarized
neutron and unpolarized proton and electron by Gardner and Zang
\cite{Gardner2001} and Garner and Plaster \cite{ Gardner2013}, and in
the neutron $\beta^-$--decay with polarized neutron and electron and
unpolarized proton by Ivanov {\it et al.}  \cite{Ivanov2018a}.

Finalizing our discussion we would like to emphasize that perspectives
of development of investigations of the neutron $\beta^-$--decays with
the contribution of the Fierz interference term $b = - 1.44\times
10^{-2}$ become real only in case of discovery of the neutron dark
matter decay modes $n \to \chi + anything$ in terrestrial laboratories
by measuring the differential cross sections for the inelastic
low--energy electron--neutron scattering $e^- + n \to \chi + e^-$ and
for the electrodisintegration of the deuteron into dark matter
fermions and protons $e^- + d \to \chi + p + e^-$ close to threshold.
These processes are induced by the same interaction $n\chi e^-e^+$
having the strength of the interaction $n \chi \nu_e \bar{\nu}_e$
responsible for the dark matter decay mode $n \to \chi + \nu_e +
\bar{\nu}_e$, allowing to explain the neutron lifetime anomaly. Of
course, indirect confirmations of existence of the neutron dark matter
decay mode $n \to \chi + \nu_e + \bar{\nu}_e$ through the evolution of
neutron stars and neutron star cooling should also testify a revision
of the experimental data on the neutron $\beta^-$--decays with
substantially improved accuracies of the measurements of correlation
coefficients and asymmetries allowing to feel the contribution of the
tensor interactions with the tensor coupling constant $C_T = -
\bar{C}_T = 1.11\times 10^{-2}$, giving the contributions of order
$10^{-4}$ to the correlation coefficients, and the Fierz interference
term $b = - 1.44\times 10^{-2}$. Such a value of the Fierz
interference term is required by the necessity to fit the rate of
neutron decay modes $n \to p + anything$, measured in beam
experiments. It is obvious that in this connection a relative accuracy
of measurements of the rate of the neutron decay modes $n \to p +
anything$ (or the neutron lifetime $\tau_n = 888.0(2.0)\,{\rm s}$) in
beam experiments should be improved up to a few parts of $10^{-4}$ or
even better.

\section{Acknowledgements}

We are grateful to Hartmut Abele for fruitful discussions stimulating
the work under this paper and to Martin Gonz\'alez--Alonso for his
comments on the value of the scalar coupling constant. The work of
A. N. Ivanov was supported by the Austrian ``Fonds zur F\"orderung der
Wissenschaftlichen Forschung'' (FWF) under contracts P31702-N27,
P26781-N20 and P26636-N20 and ``Deutsche F\"orderungsgemeinschaft''
(DFG) AB 128/5-2.  The work of R. H\"ollwieser was supported by the
Deutsche Forschungsgemeinschaft in the SFB/TR 55.  The work of
M. Wellenzohn was supported by the MA 23 (FH-Call 16) under the
project ``Photonik - Stiftungsprofessur f\"ur Lehre''.

\end{document}